# A Multicriteria Decision Making Approach to Study Barriers to the Adoption of Autonomous Vehicles

Alok Raj

Department of Production, Operations and Decision Sciences,

XLRI Xavier School of Management, Jamshedpur, India,

Email: alokraj@xlri.ac.in

J Ajith Kumar

Department of Production, Operations and Decision Sciences,

XLRI Xavier School of Management, Jamshedpur, India,

Email: akm@xlri.ac.in

Prateek Bansal

**(Corresponding Author)**

School of Civil and Environmental Engineering,

Cornell University, 301 Hollister Hall, Ithaca, NY 14853, USA

Email: pb422@cornell.edu

**Abstract**

Automation technology is emerging, but the adoption rate of autonomous vehicles (AV) will depend upon how policymakers and the government address various challenges such as public acceptance and infrastructure development. This study follows a five-stage method to understand these barriers to AV adoption. First, based on a literature review followed by discussions with experts, ten barriers are identified. Second, the opinions of eighteen experts from industry and academia regarding inter-relations between these barriers are recorded. Third, a multicriteria decision making technique, the grey-based Decision-making Trial and Evaluation Laboratory (Grey-DEMATEL), is applied to characterize the structure of relationships between the barriers. Fourth, robustness of the results is tested using sensitivity analysis. Fifth, the key results are depicted in a causal loop diagram (CLD), a systems thinking approach, to comprehend cause-and-effect relationships between the barriers. The results indicate that the lack of customer acceptance (LCA) is the most prominent barrier, the one which should be addressed at the highest priority. The CLD suggests that LCA can be mitigated by addressing two other prominent and more tangible barriers – lack of industry standards and the absence of regulations and certifications. The study's contribution lies in demonstrating that the barriers to AV adoption do not exist in isolation but are linked with each other in overlapping loops of cause and effect relationships. These insights can help different stakeholders in prioritizing their endeavors to expedite AV adoption. From the methodological perspective, this is the first study in the transportation literature that integrates Grey-DEMATEL with systems thinking.

**Keywords**: Autonomous Vehicle; Barriers; Grey-DEMATEL; Causal Loop Diagram

# 1. Introduction

In recent times, autonomous vehicles (AVs) have drawn the attention of policymakers, manufacturers, consumers, and non-governmental organizations (NGOs). AVs can revolutionize the way we travel because of their ability to move without human drivers (Gartner, 2019; MIT Technology Review Insights, 2018). As per one estimate, 30% and 50% of the US vehicle fleet will have Level 4 automation in 2040 and 2050, respectively (Litman, 2015). The logistics sector is likely to leverage full automation in the next decade. According to Chottani et al. (2018), autonomous trucks are likely to roll out in four phases: Level 3 autonomy by 2020, driverless platooning on interstate highways by 2022, Level 4 autonomy by 2025, and Level 5 autonomy by 2027.

AVs have the potential to improve urban life style, reduce crashes, reduce traffic congestion, and increase the value of travel time (Chen et al. 2017; Economist 2015; Greenwald and Kornhauser 2019). The transportation sector is a prime contributor to greenhouse gas (GHG) emissions (US EPA, 2019) and AVs are expected to also help reduce such emissions under efficient road pricing (Litman, 2019). To leverage these advantages, several leading automobile companies (Waymo, Daimler-Bosch, Ford, Volkswagen, General Motors, Toyota, Audi and Mercedes-Benz) and technology giants (Apple, Google, Tesla and Uber) are pushing their manufacturing operations to make AVs viable on the roads.

However, despite the excitement surrounding these potential advantages of AVs, there is much uncertainty among practitioners and researchers about AVs' future (Bansal and Kockelman 2017). For example, during the initial transitional period when both autonomous and conventional vehicles coexist, various traffic network strategies (such as dedicated lanes) and management strategies (such as congestion pricing) might need to be developed. Similar to any other technology or innovation, there are physical (e.g., infrastructure development) and psychological barriers (e.g., public perception) to the large-scale adoption of AVs (Bagloee et al., 2016). There is a pressing need to understand such barriers to expedite the future adoption of AVs.

Several consumer-level studies have touched upon this topic (Fagnant & Kockelman, 2015; Gkartzonikas and Gkritza 2019; Haboucha et al., 2017; Sparrow and Howard 2017). These studies have laid out a foundation to understand the barriers to AV adoption. The current study identifies two areas where this understanding can be extended. First, previous studies mostly focus on pairwise relationships between two barriers at a time (such as the impact of the lack of standards on consumer acceptance). However, any barrier can potentially influence as well as be influenced by multiple other barriers. This implies that the pairwise relationships reveal the complex reality only to some extent. Second, the relationships in previous studies are mostly associational in nature, not causal. Models that identify and depict causal influences between the barriers can better inform policymaking. The current study is thus guided by the following research questions:

   a) What are the key barriers to the adoption of AVs?
   b) How do these barriers causally influence each other?
   c) How can the causal influences be depicted and analyzed?

To answer these questions, the study follows a five-stage method that combines the Grey Decision Making Trial and Evaluation Laboratory (Grey-DEMATEL) method with the Causal Loop Diagramming approach of systems thinking. This study is the first to examine barriers to AV adoption and analyze causal relationships between them. The contribution of this study is three-fold. First, it identifies the key barriers to AV adoption and suggests a method to rank them. Second, it considers all the barriers simultaneously and elicits 'causal' relationships between them. Third, it demonstrates how Grey-DEMATEL can be integrated with systems thinking to perform structural modeling, which is the first such application in the transportation literature.

The rest of the paper is organized as follows: Section 2 presents a review of the literature and describes how the barriers to AV adoption were identified in this study; Section 3 describes how Grey-DEMATEL combined with causal loop diagramming was applied to analyze these barriers; Section 4 presents the sensitivity analyses; Section 5 discusses the key findings; and finally, Section 6 presents the conclusions and avenues for future research.

## 2. Literature review

The Society of Automotive Engineers (SAE) first formulated the definition of the AV, which was later accepted by the U.S. Department of Transportation and the National Highway Transportation Safety Administration (NHTSA, Dyble, 2018). The SAE recognizes six levels of automation in AVs starting from no automation (level 0) to full automation (level 5). In general, AVs at level 4 and above are called self-driving vehicles.

AVs, particularly those above level 4, have been a subject of discussion in recent times because of their potential to change the way we travel. Recent reviews (Gkartzonikas and Gkritza 2019; Gandia et al. 2019) suggest that AV research has grown rapidly after 2014. Researchers have mainly focused on the following themes:

a. opportunities and challenges to expect when AVs become a common mode of transport (Bagloee et al., 2016; Fagnant and Kockelman, 2015; Litman, 2018, Shladover and Nowakowski 2017; Simoni et al., 2019)

b. consumers' willingness to pay to use AVs, travel behavior, and risk perception (Buckley et al., 2018, Bansal and Kockelman, 2017; Childress et al., 2015; Daziano et al., 2017; Kröger et al., 2019; Kyriakidis et al., 2015; Schoettle and Sivak, 2014; Xu and Fan 2018)

c. system-level impact of AVs such as the effect of AVs on the design of parking systems (Nourinejad et al., 2018) and on fuel consumption (Chen et al., 2017).

## 2.1 Methodologies in AV research

To understand the market penetration of AVs and their impact on travel behavior, most previous studies have relied on stated preference (SP) surveys, followed by descriptive and econometric analyses. In many SP studies, the sample is drawn from an adult (older than 18 years) population, with some also considering subject experts and vehicle owners (Gkartzonikas and Gkritza 2019). Some of these studies have used the results of econometric models in system-level simulation frameworks to forecast long-term adoption of automation technologies (Bansal and Kockelman, 2017), to quantify impacts of AVs on national fuel consumption (Chen et al., 2017) and travel behavior (Kröger et al., 2019), and to analyze long-term innovation diffusion in automation technologies (Nieuwenhuijsen et al., 2018). In a recent study, Nourinejad et al. (2018) have adopted a mixed-integer non-linear programming approach to optimally design a parking facility for AVs.

## 2.2 Expected benefits of AVs

Several studies have briefly discussed the benefits of AVs. These include reduced transportation cost (Bagloee et al., 2016), decreased crashes (Kyriakidis et al., 2015; Li et al., 2018), reduced fuel consumption (Kyriakidis et al., 2015; Li et al. 2018), lowered traffic congestion (Fraedrich et al., 2018; Li et al., 2018), lowered driving stress (Buckley et al., 2018), enhanced critical mobility for elderly and disabled people (Litman, 2019), reduced vehicle ownership (Bagloee et al., 2016), easened parking (Nourinejad et al., 2018) and more efficient and smooth traffic circulation (Bagloee et al., 2016). While some benefits – such as relieving driving stress and easened parking – are easily acceptable, others are debatable. For example, though AVs are likely to reduce crashes and emissions per mile, induced travel demand (due to increased ease of travel) can compensate for and nullify them. Such arguments foster a sense of uncertainty in the expected benefits of AVs and, more generally speaking, point to the presence of obstacles or potential barriers that need to be addressed to accelerate AV adoption (Gkartzonikas and Gkritza 2019). Table 1 presents a summary of recent research on AVs.

**2.3  Potential barriers to AV adoption**

Even though several studies have mentioned barriers to AV adoption (Buckley et al., 2018; Fagnant and Kockelman, 2015; Kyriakidis et al., 2015; Li et al., 2018; Litman, 2019; Schoettle and Sivak, 2014), a comprehensive study that investigates the multitude of relationships between AV adoption barriers is lacking. To bridge this gap, the current study focuses on eliciting relationships between potential barriers to AV adoption.

In this study, the barriers were identified using a three-step procedure. The first step involved searching through peer-reviewed research articles listed on Scopus using the keywords-"TITLE-ABS-KEY ("autonomous vehicle") OR TITLE-ABS-KEY ("driverless cars") AND TITLE-ABS-KEY (barriers)".This initial search resulted in 87 articles. Based on the relevance criterion, 42 papers were filtered for a detailed review in the second step. Several barriers were common across these papers. Ten distinct barriers were identified by taking a union of these barriers. In the third step, six experts from industry and academia were selected based on purposive sampling and were provided with the list of barriers to check for correctness and completeness. All these experts held doctoral degrees and were knowledgeable about AVs. Detailed discussions were held individually with these experts through Skype or email. All experts agreed upon the originally identified ten barriers after minor amendments in the description. They did not suggest adding any other barrier to the list. This list of ten barriers is presented in Table 2, along with brief descriptions and literature references.

**Table 1:** Recent studies on AVs

| Authors | Focus of the study | Methodology | Opportunities | Barriers |
|---|---|---|---|---|
| Bagloee et al. (2016) | Investigate the challenges and opportunities pertaining to transportation policies that arise as a result of autonomous vehicle (AV) | Linear Programming | Reduce transportation cost, increase accessibility to low-income households and persons with mobility issues, reduction in vehicle ownership, more efficient and smooth traffic circulation | Integration of several intelligent vehicles, regulations |
| Bansal and Kockelman (2017) | Forecasting Americans' long-term adoption of connected and autonomous vehicle technologies based on policy promotion in willingness to pay | Simulation | Not mentioned | Willingness to pay |
| Buckley et al. (2018) | Drivers' responses to the experience of AVs | Simulator based experiments | Reduce stress for the drivers | Hacking and privacy |
| Chen et al. (2017) | Quantifying impacts of autonomous vehicles on national fuel consumption | Simulations | Fuel savings, traffic patterns, vehicle ownership, and land use | Not mentioned |
| Daziano et al. (2017) | Willingness to pay for the AV's | Logit model | Not mentioned | Equipment or system failure |
| Fagnant and Kockelman (2015) | Opportunities, barriers, and policy recommendations | Case study | Crash savings, travel time reduction, fuel efficiency and parking benefits | Standards for liability, security, and data privacy |
| Fraedrich et al. (2018) | Impacts of AV on built environment in the context of infrastructure | Literature, quantitative online survey, and qualitative interviews | Safety, congestion, reduction in the emission and space parking | Compatibility of AV with existing transport facilities, infrastructure planning |

| Kröger et al. (2018) | Impact of AV on travel behaviour for Germany and USA | Simulation | Reduction of value-of-travel-time-savings | Not mentioned |
|---|---|---|---|---|
| Kyriakidis et al. (2015) | User acceptance, concerns, and willingness to buy partially, highly, and fully automated vehicles | Survey | Traffic crashes, reduction in pollution | Hacking and privacy, legal issues and safety |
| Li et al. (2018) | Analyzes the emerging importance and research frontiers in formulating highly AV policies | Literature review | Lowering emissions, providing critical mobility to the elderly and disabled, expanding road capacity, reducing mortality | Government regulations, licensing and testing standards, certification, reliability, legal challenges |
| Litman (2018) | Explores autonomous vehicle benefits and costs, and impacts on transportation planning issues | Literature review and experts opinion | Reduced traffic and parking congestion, independent mobility for low-income people, increased safety, energy conservation and pollution reduction | Social equity concerns, reduced employment, increased infrastructure costs, reduced security, hacking and privacy |
| Nourinejad et al. (2018) | Autonomous vehicles will have a major impact on parking facility designs in the future | Mixed-integer non-linear program | Space utilization | Not mentioned |
| Schoettle and Sivak (2014) | A survey of public opinion about autonomous and self-driving vehicles in the U.S., the U.K., and Australia | Survey | Fewer crashes, less traffic congestion, shorter travel time, lower vehicle emissions | Security issues, data privacy, interacting with non-self-driving, safety concerns of equipment failure |
| Shladover and Nowakowski (2017) | Regulatory challenges for road vehicle automation under the context of California | Survey | Transportation system performance and safety | Absence of clearly defined standards and testing procedures |
| Xu and Fan (2018) | Risk perceptions and anticipation of insurance demand for autonomous vehicles in the Chinese market. | Survey | Not mentioned | Operating error risk |

**Table 2.** Barriers to the adoption of autonomous vehicles

| S No. | Barriers | Code | Implied Meaning | References |
|---|---|---|---|---|
| 1 | Reduced security and privacy | RSP | AVs are likely to store a large amount of personal data (such as trip patterns and users preferences) and may be vulnerable to leakage of such information. | Fagnant and Kockelman (2015); Clark et al. (2016); Litman (2019); Schoettle and Sivak, (2014); Buckley et al. (2018); Kyriakidis et al. (2015); Sheehan et al. (2018) |
| 2 | Social inequity | SIN | Initial cost of AVs is likely to be much higher when compared to their counterpart driver-operated vehicles. Thus, only wealthy consumers might be able to afford AVs as personal vehicles. | Cohen (2016); The Economist (2018); Litman (2019); Tech Policy Lab (2017) |
| 3 | Obscurity in accountability | OSA | OSA refers to the lack of clarity in identifying who is accountable for the accidents and/or damages related to AVs- the owner, the manufacturer, or someone else? | Fagnant and Kockelman (2015); Li et al. (2018) Soble and Lucia (2015) ; J D Power (2018) |
| 4 | Lack of customer acceptance | LCA | If potential customers do not accept the AV as an alternative to manned vehicles and do not show confidence in it, the adoption of AVs cannot be expedited. | Bagloee et al. (2016); Buckley et al., 2018); Li et al. (2018); The Economist (2018); The Gartner (2017); Gramlich, (2018) |
| 5 | Potential loss of employment | PLE | AVs will replace human drivers and can have a significant impact on employment. This can be a barrier to the popularity and subsequent growth of AVs. | Litman (2019); Balakrishnan (2017); O'Brien (2017) |
| 6 | Inadequate infrastructure | INF | Huge infrastructure investments are required to make AVs viable on the road. AVs might need a dedicated lane, which requires additional investment. Deployment of smart technologies is essential to enable vehicle to vehicle (V2V) and vehicle to infrastructure (V2I) communications. | Clark et al. (2016); Fraedrich et al. (2018); Fagnant and Kockelman (2015) |
| 7 | Lack of standards | LOS | AVs are likely to be operated on a network, wherein they can talk and respond to each other to avoid crashes and escape traffic jams. For this purpose, AVs manufactured by different companies must follow standards so that they can fully leverage the advantages of automation through efficient communication. However, making this happen, particularly in emerging markets, has its challenges. | Fagnant and Kockelman (2015); Smith (2018) |

| # | Barrier | Code | Description | References |
|---|---------|------|-------------|------------|
| 8 | Absence of regulation and certification | ARC | There is a lack of consistent certification framework and standardized set of safety norms for the acceptance across different levels. Under these circumstances, AV manufacturers and suppliers may encounter regulatory uncertainty, leading to slower technological innovations. | Fagnant and Kockelman (2015); Bansal and Kockelman (2017); Li et al. (2018) Shladover and Nowakowski (2017); NCSL (2018) |
| 9 | Manufacturing cost | MNC | The high manufacturing cost of AVs is one of the key barriers to their adoption on a mass scale. | Fagnant and Kockelman (2015); Bansal and Kockelman (2017); Shchetko (2014); David and Elisabeth (2018) |
| 10 | Induced travel | ITRL | AVs could increase the vehicle miles traveled and urban sprawl. AVs is likely to reduce travel times and emissions, such savings can be offset by an increase in the demand for travel. Vehicle miles traveled can also be induced due to shift from public transit to low-occupancy AVs. | Bansal et al. (2016) ; Haboucha et al. (2017); Truong et al. (2017); Gkartzonikas and Gkritza (2019) |

## 3. Research methodology

The study follows a five-stage methodology. First, a set of key barriers is identified based on a literature review and discussions with experts. Second, a survey of experts from academia and industry is conducted to gather pertinent data on how mitigating a given barrier would affect other barriers. Third, Grey-DEMATEL is applied on this data a) to rank the barriers and b) to segregate them into cause and effect categories. DEMATEL is a well-known method in the discipline of multi-criteria decision-making (Si et al., 2018). Fourth, a sensitivity analysis is conducted on these results using different expert weighting schemes to check their robustness. Fifth, the results of the Grey-DEMATEL model are depicted and analyzed in a causal loop diagram, a systems thinking technique, to help prioritize the barrier-mitigation policies for the mass adoption of AVs. Figure 1 shows a schematic flow-chart of these stages.

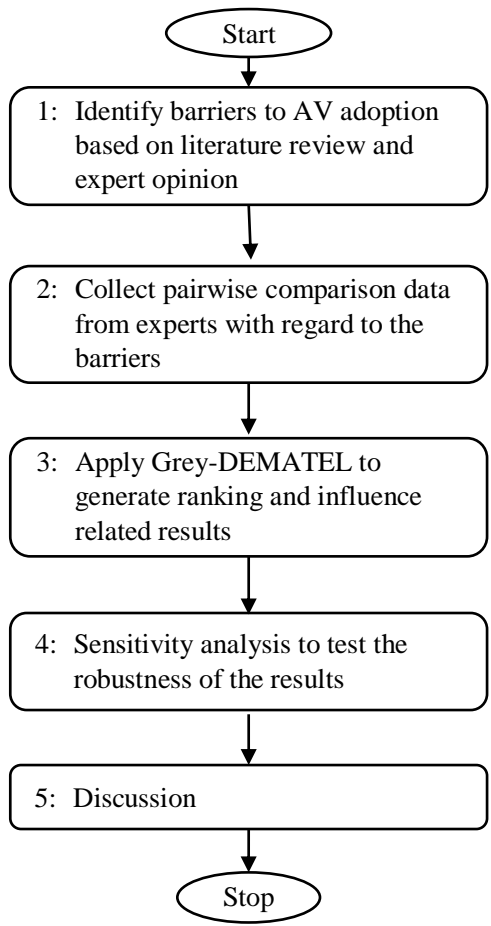

**Figure 1.** The stages in the study

The US was chosen as the geographical context of the study, for two reasons. First, the US was amongst the global leaders in AV innovation and development, but experienced challenges to AV adoption. Second, in 2018, the US ranked third globally in terms of the "Autonomous Vehicle Readiness Index", but slipped to fourth place in 2019 (KPMG, 2019), suggesting the presence of specific barriers that might have played a role in slowing down the progress of AVs in the country.

## 3.1 The choice of Grey-DEMATEL

Some of the widely used MCDM methods are Analytic Hierarchy Process (AHP), Interpretive Structural Modeling (ISM), Analytic Network Process (ANP) and DEMATEL (Acuña-Carvajal et al., 2019; Luthra et al. 2018). AHP can only be used to derive rankings of factors, while ISM helps evaluate contextual relationships between them. ANP evaluates ranking, assists in revealing interdependencies between the factors and manages the issue of consistency as well. However, ANP has limited applicability due to its complex procedure (Luthra et al. 2018; Mangla et al., 2018). DEMATEL goes beyond ANP and helps separate the constituents of a system into cause and effect groups. The advantages of DEMATEL over AHP, ISM, and ANP are well-established in the literature (Acuña-Carvajal et al., 2019; Gölcük and Baykasoğlu 2016; Luthra et al. 2018; Mangla et al., 2018). Previous studies have applied DEMATEL in diverse fields such as transportation service quality (Liou et al., 2014), recycling of e-waste (Rahman and Subramanian, 2012), supplier selection (Govindan et al., 2018), third-party logistics (Govindan and Chaudhury, 2016), and the selection of renewable energy resources (Buyukozkan and Guleryuz, 2016). A review of a pool of 346 papers by Si et al (2018) provides a good account of the use of DEMATEL in engineering and management research.

DEMATEL relies on the subjective opinions of experts. Since using subjective opinions can potentially infuse uncertainty and bias in the input data, DEMATEL is sometimes used in conjunction with grey system theory. Grey theory has the ability to generate satisfactory results when the available data is somewhat limited or incomplete, or when the uncertainty and variability in the factors is high (Bai and Sarkis, 2013). Previous studies also note that grey theory can enhance the exactness of human judgments when integrated with the decision-making process (Bai and Sarkis, 2010, 2013; Tseng, 2009). Examples of research using Grey-DEMATEL include analyzing the enablers of risk mitigation in electronic supply chains (Rajesh and Ravi 2015), the risk faced by third-party logistics service providers (Govindan and Chaudhuri 2016), the critical factors of green business failure (Cui et al. 2018) and the barriers to the adoption of environmentally friendly products (Shao et al., 2016). The elements of Grey-DEMATEL in this study have been adapted from Bai and Sarkis (2010), Govindan and Chaudhuri (2016) and Rajesh and Ravi (2015). In addition to the method followed by these researchers, Causal Loop Diagramming (CLD), a systems thinking approach, has been used to better comprehend causal relationships. In all, the method involves 10 steps, which are explained in Appendix 1.

## 3.2 Applying Grey-DEMATEL

In step 1, experts in academia and industry, who hold at least a Master's degree in transportation engineering or planning and have published research papers or reports pertaining to AVs were identified following a purposive sampling approach. In all, 55 experts were contacted in the US via email between October 2018 and December 2018, and 18 completed responses were received. Of these, 14 were from academia and 4 were from the industry. A majority of the experts (14 of 18) had worked in the field of AVs for more than 3.5 years, while all the academics held at least a doctoral degree. Table 3 shows the affiliations and qualifications of the experts who responded to the survey. The survey was presented using two Excel sheets. The first sheet described the barriers and the second, solicited experts' opinions about the extent of influence of each barrier on the other nine barriers, using a linguistic scale ("No" to "Very High", see Table 4). These are also known as pairwise comparisons. By this, 18 direct-relation matrices, each of size *10* x *10*, were obtained.

Steps 2 – 8 were followed as described in Appendix 1. Step 8 yields an *R* and a *C* value for each barrier. *R* represents the total influence that a given barrier has on other barriers, while *C* represents the total influence that other barriers have on the given barrier. From them, *R+C* and *R–C* values are computed for each barrier. The *R+C* value indicates the prominence of the barrier within the system of barriers, since a high *R+C* means that a barrier simultaneously has a large influence on the other barriers *and* is influenced highly by them, while a low *R+C* suggests that both types of influence are low. The *R–C* value stands for the net influence of a barrier since it is the difference between how much a barrier influences other barriers and how much it is influenced by them. More specifically, the *R–C* score indicates the barrier's propensity to be either a cause (influencer / driver) or an effect (influenced / receiver) in relation to other barriers in the system. If it is positive, the barrier is likely to be a "cause barrier", one that influences other barriers more than being influenced by them. If *R–C* is negative, then it is taken to be an "effect barrier", or one that is influenced more by others than influencing them. Thus, the sign of *R–C* helps in classifying the set of barriers into two groups

– "cause" and "effect". See Table 5 for the *R*, *C*, *R+C* and *R–C* values for the barriers in the current study, as well as their respective rankings on *R+C* and *R–C*. Table 5 helps identify the cause, effect and prominence barriers as well. Following Step 9, the Influence-Prominence Map (IPM) was plotted as shown in Figure 2.

**Table 3:** Affiliations and qualifications of the experts

| S. No. | Affiliation | Type | Qualification |
|---|---|---|---|
| 1 | Department of Civil and Materials Engineering, University of Illinois at Chicago, USA | Academic | PhD |
| 2 | Department of Civil and Environmental Engineering, University of Michigan, USA | Academic | PhD |
| 3 | Autonomous Systems Laboratory, Stanford University | Academic | PhD |
| 4 | Florida Atlantic University | Academic | PhD |
| 5 | Department of Civil and Materials Engineering, University of Illinois at Chicago, USA | Academic | PhD |
| 6 | Institute of Transportation Studies, University of California Davis | Academic | PhD |
| 7 | Centre for Urban Transportation Research - University of South Florida | Academic | PhD |
| 8 | University of Texas at Austin | Academic | PhD |
| 9 | Department of Civil and Environmental Engineering, University of Illinois at Urbana-Champaign, USA | Academic | PhD |
| 10 | Cornell University | Academic | PhD |
| 11 | Princeton University | Academic | PhD |
| 12 | Centre for Sustainable Systems, University of Michigan | Academic | PhD |
| 13 | Michigan State University | Academic | PhD |
| 14 | Department of Civil and Materials Engineering, University of Illinois at Chicago, USA | Academic | PhD |
| 15 | Active Transportation, Transpo Group | Practitioner | PhD |
| 16 | Kettering University | Practitioner | PhD |
| 17 | Senior Modeller at Puget Sound Regional Council | Practitioner | Masters |
| 18 | United States Environmental Protection Agency, EPA | Practitioner | Masters |

**Table 4:** Grey values for the linguistic scale used for expert assessments.

| Linguistic terms | Grey values |
|---|---|
| No influence (N) | [0, 0] |
| Very low influence (VL) | [0, 1] |
| Low influence (L) | [1, 2] |
| Medium influence (M) | [2, 3] |
| High influence (H) | [3, 4] |
| Very high influence (VH) | [4, 5] |

**Table 5:** Degree of prominence and net cause/effect values

| Barriers | R | C | R+C | R – C | Rank as per R+C (Prominence) | Rank as per R-C (Net Influence) | Cause / Effect |
|---|---|---|---|---|---|---|---|
| RSP | 0.315 | 0.275 | 0.590 | 0.040 | 7 | 5 | C |
| SIN | 0.213 | 0.281 | 0.494 | -0.068 | 9 | 8 | E |
| OSA | 0.404 | 0.311 | 0.714 | 0.093 | 4 | 3 | C |
| LCA | 0.483 | 0.692 | 1.175 | -0.209 | 1 | 10 | E |
| PLE | 0.068 | 0.063 | 0.132 | 0.005 | 10 | 6 | C |
| INF | 0.398 | 0.310 | 0.708 | 0.088 | 5 | 4 | C |
| LOS | 0.501 | 0.399 | 0.900 | 0.102 | 2 | 2 | C |
| ARC | 0.419 | 0.431 | 0.850 | -0.012 | 3 | 7 | E |
| MNC | 0.415 | 0.281 | 0.695 | 0.134 | 6 | 1 | C |
| ITRL | 0.184 | 0.357 | 0.541 | -0.173 | 8 | 9 | E |

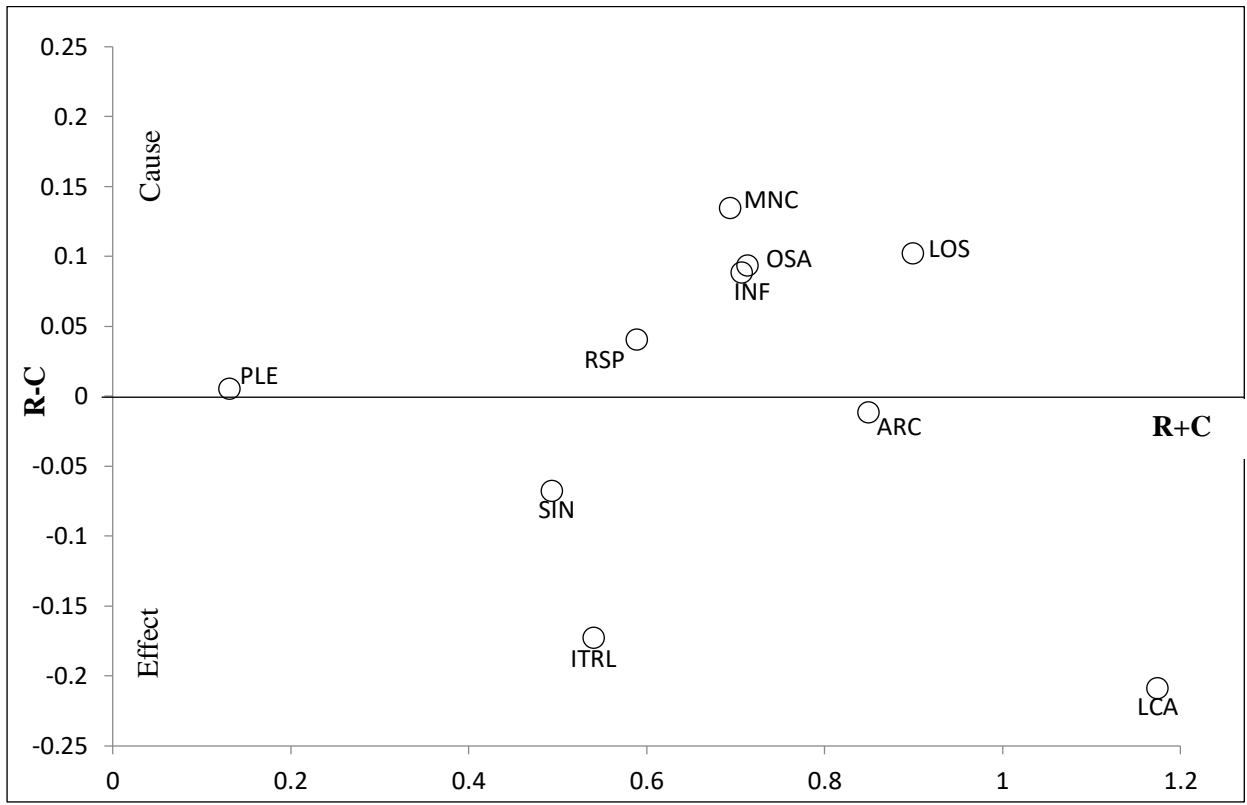

**Figure 2**. Influence Prominence Map

### 3.3 Causal Loop Diagram

Traditionally, DEMATEL also involves plotting the causal relationships between the factors on the IPM using arrows. In the current study, a Causal Loop Diagram (CLD, see Step 10 of Appendix 1) has been used to the depict the influences (Figure 3), instead of the IPM, as it provides a more elegant and effective way to represent and comprehend the causal influences between entities in a complex system. The threshold in the current study was set as $\theta = \mu + \sigma$, which evaluates to: $0.0375 + 0.0289 = 0.0665$. This led to the identification of 17 above-threshold influences forming 10 feedback loops as shown in the CLD in Figure 3. It should be noted here that in cases when both $m_{ij}$ and $m_{ji}$ are at least $\theta$ (meaning that both factors $i$ and $j$ influence each other prominently), there are two arrows linking factors $i$ and $j$ in opposite directions, resulting in feedback loops that involve only two barriers.

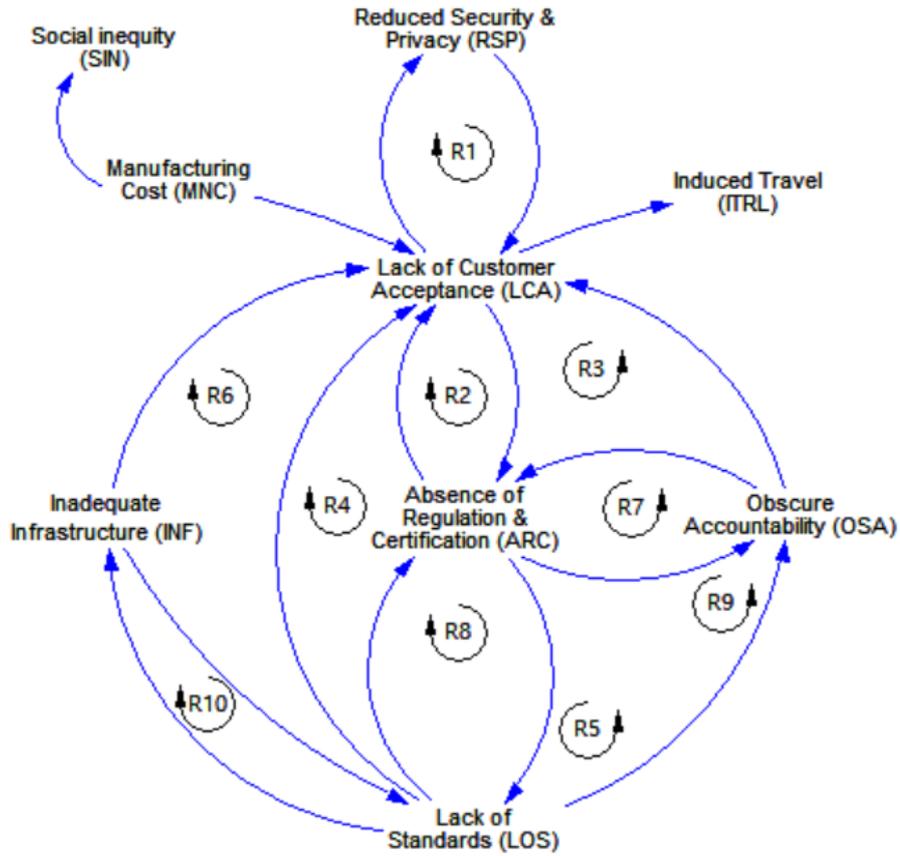

**Figure 3.** Causal loop diagram

## 4. Sensitivity Analysis

In comparison to past DEMATEL-based research that has found causal relationships using data gathered from seven or fewer experts (Cui et al. 2018; Bai and Sarkis 2010; Awasthi et al., 2018), the current study had a larger sample, constituted by eighteen experts. Despite this, the assignment of equal weightages to the experts despite differences in their experience can question the robustness of the results. To test robustness, a sensitivity analysis was carried out. The experts were divided into three groups on the basis of their experience – more than 8 years, 5 to 8 years, and 3.5 to 5 years – and different weights were assigned to respective groups to create six alternative scenarios. For example, in the first scenario, 50%, 30%, and 20% weight were assigned to experts with the experience of more than 8 years, 5 to 8 years, and 3.5 to 5 years, respectively. The Grey-DEMATEL method was applied in each of these scenarios with the same pairwise comparison data gathered from the experts, with the intention to examine how three key outcomes change with respect to the base scenario: 1) the barrriers' ranks on R+C, 2) their ranks on R–C and 3) the inter-barrier influences that fall above the threshold $\theta = \mu + \sigma$. Table 6 shows that across six scenarios, the R+C rank changes atmost by 3 for one barrier (for ITRL), by 2 for four of the barriers (RSP, SIN, LOS and ARC), by 1 for two barriers (OSA and INF) and does not change at all for two barriers (LCA, PLE and MNC). Likewise, Table 7 shows that the R–C rank changes atmost by 2 for five of the barriers (RSP, SIN, INF, LOS and ITRL), by 1 for OSA and ARC and does not change for the same three barriers (LCA, PLE and MNC). The relatively low rank changes across scenarios suggests that the ranks obtained in the base scenario are fairly robust.

**Table 6:** Sensitivity analysis for degree of prominence

| Barriers code | Rank as per R+C (degree of prominence) | | | | | | | |
|---|---|---|---|---|---|---|---|---|
| | Rank in Base Scenario | Rank in Scenario 1 | Rank in Scenario 2 | Rank in Scenario 3 | Rank in Scenario 4 | Rank in Scenario 5 | Rank in Scenario 6 | Maximum change in rank |
| RSP | 7 | 6 | 6 | 7 | 7 | 6 | 5 | 2 |
| SIN | 9 | 9 | 9 | 8 | 7 | 9 | 9 | 2 |
| OSA | 4 | 3 | 3 | 4 | 4 | 4 | 3 | 1 |
| LCA | 1 | 1 | 1 | 1 | 1 | 1 | 1 | 0 |
| PLE | 10 | 10 | 10 | 10 | 10 | 10 | 10 | 0 |
| INF | 5 | 4 | 4 | 4 | 5 | 5 | 5 | 1 |
| LOS | 2 | 2 | 2 | 2 | 3 | 1 | 2 | 2 |
| ARC | 3 | 3 | 3 | 3 | 2 | 2 | 1 | 2 |
| MNC | 6 | 6 | 6 | 6 | 6 | 6 | 6 | 0 |
| ITRL | 8 | 7 | 9 | 8 | 6 | 7 | 8 | 3 |

Table 8 shows that the number of inter-barrier influences that are above the threshold, which is 17 in the base scenario, varies a little across the six alternate scenarios. It remains 17 in three of them (Scenarios 1, 5 and 6) but becomes 18, 19 and 20 in Scenarios 3, 4 and 2 respectively. Out of 17 base scenario influences, nine are present in all the six alternate scenarios, three appear in five of the alternate scenarios and four appear in four of the alternate scenarios. Thus, 9+3+4 = 16 of the 17 base scenario influences appear in at least four of the six alternate scenarios, while the remaining one (ARC-OSA) appears in only two of them. This implies that the CLD drawn for the base scenario will overlap considerably with the CLDs of the alternate scenarios, indicating that the CLD and the set of causal relationships included in it are reasonably robust.

**Table 7:** Sensitivity analysis for net influence

| Barriers code | Rank as per R-C | | | | | | | |
|---|---|---|---|---|---|---|---|---|
| | Rank in Base Scenario | Rank in Scenario 1 | Rank in Scenario 2 | Rank in Scenario 3 | Rank in Scenario 4 | Rank in Scenario 5 | Rank in Scenario 6 | Maximum change in rank |
| RSP | 5 | 6 | 4 | 4 | 5 | 5 | 5 | 2 |
| SIN | 8 | 6 | 7 | 8 | 6 | 8 | 8 | 2 |
| OSA | 3 | 3 | 3 | 2 | 3 | 3 | 2 | 1 |
| LCA | 10 | 10 | 10 | 10 | 10 | 10 | 10 | 0 |
| PLE | 6 | 6 | 6 | 6 | 6 | 6 | 6 | 0 |
| INF | 4 | 3 | 4 | 5 | 5 | 5 | 4 | 2 |
| LOS | 2 | 3 | 2 | 1 | 2 | 2 | 2 | 2 |
| ARC | 6 | 5 | 6 | 5 | 6 | 6 | 6 | 1 |
| MNC | 1 | 1 | 1 | 1 | 1 | 1 | 1 | 0 |
| ITRL | 9 | 8 | 8 | 8 | 7 | 7 | 9 | 2 |

**Table 8:** Sensitivity analysis for number of inter-barrier influences

|  | **Base Scenario** | **Scenario 1** | **Scenario 2** | **Scenario 3** | **Scenario 4** | **Scenario 5** | **Scenario 6** |
|---|---|---|---|---|---|---|---|
| μ + σ | 0.0666 | 0.0723 | 0.0520 | 0.0623 | 0.0589 | 0.0670 | 0.0750 |
| No. of relationships | 17 | 17 | 20 | 18 | 19 | 17 | 17 |
| RSP - LCA | 1 | 1 | 1 | 1 | 1 | 1 | 1 |
| OSA - LCA | 1 | 1 | 1 | 1 | 1 | 1 | 1 |
| OSA -ARC | 1 | 1 | 1 | 1 | 1 | 1 | 1 |
| LCA - RSP | 1 | 0 | 1 | 1 | 1 | 1 | 1 |
| LCA- ARC | 1 | 1 | 1 | 1 | 1 | 1 | 1 |
| LCA- ITRL | 1 | 1 | 0 | 1 | 0 | 1 | 1 |
| INF-LCA | 1 | 1 | 1 | 1 | 1 | 1 | 1 |
| INF-LOS | 1 | 1 | 0 | 1 | 0 | 1 | 1 |
| LOS-OSA | 1 | 1 | 0 | 1 | 0 | 1 | 1 |
| LOS-LCA | 1 | 0 | 1 | 1 | 1 | 1 | 0 |
| LOS-INF | 1 | 1 | 1 | 1 | 1 | 1 | 1 |
| LOS-ARC | 1 | 1 | 1 | 1 | 1 | 1 | 1 |
| ARC-OSA | 1 | 0 | 0 | 1 | 0 | 1 | 0 |
| ARC-LCA | 1 | 0 | 1 | 1 | 1 | 1 | 1 |
| ARC-LOS | 1 | 1 | 1 | 1 | 1 | 1 | 1 |
| MNC-SIN | 1 | 1 | 1 | 1 | 1 | 1 | 1 |
| MNC-LCA | 1 | 1 | 1 | 1 | 1 | 1 | 0 |
| RSP - LOS | 0 | 0 | 1 | 0 | 1 | 0 | 0 |
| SIN-MNC | 0 | 0 | 1 | 0 | 1 | 0 | 0 |
| LCA- INF | 0 | 0 | 1 | 0 | 1 | 0 | 0 |
| LCA- MNC | 0 | 0 | 1 | 1 | 1 | 0 | 0 |
| INF-ARC | 0 | 1 | 0 | 0 | 0 | 0 | 1 |
| INF-ITRL | 0 | 1 | 1 | 0 | 1 | 0 | 1 |
| ARC-INF | 0 | 0 | 1 | 0 | 1 | 0 | 1 |
| ITRL-LCA | 0 | 0 | 1 | 0 | 0 | 0 | 0 |

**Note:** 1 indicates relationship exist between two barriers while 0 indicates otherwise

## 5. Discussion

In this section, the extents of prominence and net influence of the barriers to AV adoption are discussed using *R-C* scores, *R+C* scores (Table 5), and the causal loop diagram (Figure 3).

### 5.1 R–C and R+C scores

The *R-C* scores in Table 5 suggest that MNC, LOS, OSA, INF, RSP and PLE, can be considered as cause factors (in decreasing order of net outward influence) while LCA, ITRL, SIN and ARC as effect factors (in decreasing order of net inward influence).

LCA, the lack of customer acceptance, is ranked 10 on *R–C*, indicating it has the greatest net inward influence amongst the barriers. Interestingly, it is also ranked 1 on the *R+C* score, which means that it also has the highest prominence in the system of AV barriers. The prominence of LCA suggested by *R+C* is also consistent with LCA's position in the CLD (Figure 3). Six of 10 feedback loops in the CLD involve LCA and are labelled R1 through R6. The other four loops labelled R7 through R10 have variables in common with, and are linked to, these first six loops. LCA influences three barriers prominently and is influenced by six barriers, that is, it is involved in nine causal relationships in the CLD, which is the most for any barrier in the study. This suggests that LCA plays a fundamental role in the adoption of AVs. The KPMG report (2019) and several academic studies also indicate that LCA is a major challenge in the adoption of AVs (Daziano et al. 2017; Xu et al. 2018; Threlfall 2018; Bansal and Kockelman 2017; Haboucha et al. 2017). This is further corroborated by the American Automobile Association report (2017), which reveals that 78% of Americans have fear of riding AVs. A more recent study carried out in the European Union suggests that people are uncomfortable towards driverless cars and trucks as well (Hudson et al., 2019). Therefore, building trust among customers and gaining their acceptance is very important for the success of AVs (Buckley et al., 2018).

Generally speaking, the more prominent barriers should be addressed first by the government, the policymakers and managers, for the faster market diffusion of AVs. After LCA, the next prominent barrier is LOS, which is followed by ARC. Whereas standardization is important to enable efficient communication among vehicles developed by different companies, certification and testing of AVs is crucial to uphold the safety of travelers and to attain industrial standards. These factors could hamper the production of AVs and lead to a mismatch between the demand and the supply of AVs in future. For instance, if customer acceptance increases in the future and AVs are seen on local streets, then more consumers may wish to own one and AV manufacturers such as Tesla may not be able to immediately produce enough to meet the demand.

OSA is the fourth prominent barrier to the adoption of AVs. In relation to this barrier, the 2014 RAND study notes that the key questions that need to be addressed include who the responsible will be in the case of an accident and how the liability will be distributed among different stakeholders (Anderson et al., 2014). To this end, leading innovators in driverless technology such as Google, Mercedes Benz, and Volvo have decided to take responsibility in the case of accident due to a technological flaw (Ballaban, 2015). However, an accident may happen due to a combination of multiple reasons and a sequence of events. Thus, more specific guidelines need to be prepared by lawmakers. Insurance companies might be afraid of participation due to high compensations in case of damages (governed by high vehicle cost) and complexities of vehicle components.

INF, inadequate infrastructure, emerged as the fifth important barrier in the analysis. Here, Fraedrich et al. (2018) suggests that consumers are skeptical about the compatibility of AVs with existing transport and urban planning objectives. For the rapid adoption of AVs, highly-maintained and well-marked roads, high density and accessibility to electric charging stations, and network infrastructure for seamless communication are essential. As per the recent KPMG report (2019), the US is not at par with other developed economies such as the Netherlands and Singapore in terms of infrastructure. The US is ranked seventh on this dimension of the Autonomous Readiness Index. Hence, government organizations need to focus on improving the infrastructure for AVs.

Manufacturing cost, ranked 1 in terms of $R$–$C$ score, seems to have the greatest <u>net outward influence</u> on other barriers to AV adoption in the system. Thus, AV adoption can be quickened if the government provides incentives to AV manufacturers to invest in research and development to make automation technology more viable. Here, a reduction in component prices will also help tremendously. For instance, among the most expensive components in AVs are the Light Detection and Ranging (LiDAR) sensors. Their unit price, which was around $70,000 in the protoyping stage, fell to around $6,000 later and may drop to $250 once companies reach mass production (IndustryWeek, 2018).

## 5.2  CLD: feedback loops

The CLD in Figure 3 complements the insights provided by the *R–C* and *R+C* scores and helps in comprehending relationships between the barriers in terms of feedback loops. Examining the six feedback loops involving LCA suggests certain patterns in the relationships. Loop R1 (LCA ↔ RSP) represents the mutual influence of LCA and RSP on each other. This is consistent with Sener et al. (2019), who found that security and privacy negatively affect Texans' intentions to use AVs.  It is also aligned with Kyriakidis et al. (2015), whose extensive study across 109 countries also indicated similar results. In Bagloee et al. (2016) too, system security and integrity emerged as a serious concern, due to which customers might be reluctant to use AVs. When customers perceive security and privacy to be wanting, their acceptance of AVs is likely to be low. In contrast, an increase in either of them can be driven by, as well as result in, an increase in the other. Recently, US governments have enacted a new legislation known as the SPY Car Act on data privacy that provides jurisdiction to the NHTSA to protect the use of driving data in all vehicles manufactured for sale in the US (Taeihagh and Lim, 2019).  This is likely to favour customers' acceptance of AVs.

LCA and ARC also mutually influence each other, as denoted by Loop R2 (LCA ↔ ARC). A US-based study conducted by The Association for Unmanned Vehicle Systems International (AUVSI) reveals that regulatory framework is a concern for AV adoption (Hyde, 2019).  In that study, 54% of the respondents preferred that AV-related regulations should come from the US Department of Transportation and not from individual states. Due to the absence of federal regulations, many states have formulated conflicting regulations related to the testing and licensing of AVs (Autonomous Vehicles Survey Report, 2019).  In the absence of a consistent regulation or framework, AV manufacturers may face uncertainties regarding testing and certification (Fagnant & Kockelman, 2015). In turn, if customer acceptance increases across the country, it can be expected that the federal government will be under more pressure to better define AV related regulations and thus regulation and certification will gain greater clarity and maturity.

Loop R2 includes the direct influence of ARC on LCA. However, ARC influences LCA indirectly as well, through its influences on other variables, and the sequential influences of those variables on yet other variables. These give rise to the four loops, R3 through R6.  In loop R3 (LCA → ARC → OSA → LCA), the influence of ARC on OSA is key. That is, the absence of regulation and certification leads to greater obscurity in accountability. It can also be reasoned that when regulations improve, rules that specify who is accountable in the event of accidents or untoward incidents will also develop and become more clear. Further, OSA influences LCA, that is, when there is not enough clarity on the liabilities related to AV, it can discourage customers from accepting AVs, thus completing loop R3. Loop R4 (LCA → ARC → LOS → LCA) is generated owing to ARC's influence on LOS. The absence of regulation and certification prevents the development of industry standards pertaining to AVs. Here, Li et al. (2018) discuss that a lack of standards can contribute to customers' hesitation to purchase AVs. Shladover and Nowakowski (2017) indicate that even safety is hard to certify by the developer in the absence of well-defined standards. Whereas previous studies provide pair-wise associations of LOS with other barriers, this study reveals more intricate causal associations between LOS and other barriers.

Both loops R5 and R6 branch out from Loop R4, at LOS. Loop R5 (LCA → ARC → LOS → OSA → LCA) and loop R6 (LCA → ARC → LOS → INF → LCA) come into existence owing to the influence of the lack of standards on the level of obscurity in accountability and on the extent of the available infrastructure pertaining to AVs, respectively. These findings suggest that the absence of country-wide standards would slow down the development of rules related to liabilities as well as the necessary physical infrastructure. The remaining four loops (R7 through R10) in the CLD are not independent of the first six loops described above (R1 through R6). Rather, they are formed owing to mutual relationships between some of the barriers. The first three of them R7 (ARC ↔ OSA), R8 (ARC ↔ LOS) and R9 (ARC → LOS → OSA → ARC) involve mutual relationships between ARC, LOS and OSA, while R10 (LOS ↔ INF) involves LOS and INF.

In sum, the lack of consumer acceptance (LCA) is the most prominent barrier to AV adoption, but associations related to AVs (e.g., government or manufacturers) should perhaps focus on mitigating more tangible barriers – the lack of standards (LOS) and the absence of regulations and certifications (ARC), which are not only ranked second and third in terms of prominence, but also significantly affect other barriers (including LCA) through various mechanisms.

To this end, the National Highway Traffic Safety Administration (NHTSA)[1] has already started to develop industry standards in the US, but is facing challenges since much of the technology is in the form of trade secrets (NHTSA, 2017). NHTSA has also outlined vehicle performance guidance for AV manufacturers (Taeihagh and Lim, 2019), which can help improve industry standards. Such guidelines are also crucial in the mass deployment of AVs since the ecosystem of AVs would become complex if different manufacturers use different protocols for their models. Standardized design and manufacturing of AVs would enable them to communicate with each other and would facilitate the improvement of the infrastructure.[2] In fact, Taeihagh and Lim (2019) note that standardization is vital from the litigation perspective – probably due to the lack of industry standards, the US federal government is not formulating nation-wide standard rules regarding the allocation of liability and insurance to the concerned party.[3]

---

[1] The NHTSA provides guidelines and regulates different entities involved in manufacturing, designing, supplying, testing, selling, operating and deploying AVs in the US.

## 5.3 Other insights

Apart from the feedback loops, the CLD also shows that manufacturing cost prominently influences customer acceptance. Intuitively, it can be reasoned that if manufacturing cost increases, then customer acceptance of AVs would decrease and vice-versa. This result corroborates with CarInsurance.com's survey in the US, which reveals that 34% and 56% respondents showed interest in buying a car with strong and moderate level of automation respectively if companies offered 80% discount on AVs (Bansal et al., 2016). Two barriers – social inequity and induced travel – also do not influence any other barrier but are each influenced by one other barrier. Social inequity is influenced by manufacturing cost. This is expected since higher manufacturing costs imply higher sticker prices for AVs, which in turn means that only a narrow segment of the society can afford AVs. Induced travel is influenced by the lack of customer acceptance. Once consumers are convinced about the benefits of AVs (e.g., reduced travel time and cost), they are likely to drive more vehicle miles.

## 6 Conclusions and future work

Autonomous vehicles (AVs) are now on the cusp of commercialisation and academic interest in AVs is growing. The current study is relevant in this backdrop as it draws attention to key barriers to AV adoption and offers insights on prioritizing the policies to overcome them. To this end, the study views barriers to AV adoption as the components of a system, which mutually influence each other. To understand this system, the study analyzes the relationships between the barriers using Grey-DEMATEL and systems thinking.

---

[2] As per a recent KPMG (2018) report, the US has relatively fewer charging stations, poorer road quality and infrastructure in comparison to The Netherlands or Singapore.
[3] Litigation over AVs is still in its infancy in the US and has not been tested in the court.

This study's results have several societal and practical implications for manufacturers, policymakers and the government. The analysis shows that a lack of consumer acceptance is the biggest barrier to AV adoption. To gain the trust of consumers, multiple stakeholders are required to work in concert. For example, government entities may need to intervene and enforce standardized AV production and testing regulations across the US. Technology innovators and manufacturers can focus on reducing costs, which will also help address social inequity concerns. Introducing AVs as a shared mode is likely to attain both objectives because vehicle costs would become irrelevant and standardization would be much easier with same-vehicle fleets. Besides, this study indicates that policymakers need not worry too much about the employment loss due to AVs, which is generally hyped as an important concern. Perhaps, the experts think that the loss in jobs due to automation is likely to be compensated for by newly created jobs.

The study has some limitations. First, it relies on the opinions of only eighteen experts and is specific to the context of the US. While a sensitivity analysis supports the robustness of the obtained results, it is possible that involving more experts and studying over a wider geographical context can reveal finer aspects. Second, new technologies generally gain consumer acceptance when their benefits are evident. However, automobiles impose external costs such traffic congestion, accident risk and pollution vis-à-vis other technologies such as smart phones, personal computers and digital cameras. This means that even if individual consumers accept AVs, there could be resistance to AVs at a collective, or community, level. Hence, community acceptance is also important to the adoption of AVs. Though community acceptance did not emerge as a barrier in the current study, it cannot be overlooked and future studies should explore the potential role that it can play in the adoption of AVs.

Despite these limitations, this study provides unique insights about the causal relationships between barriers which cannot be derived from consumer behavior studies relying on large samples. An overarching contribution of the study lies in understanding that the barriers to AV adoption <u>do not act in isolation, but as a system of interrelated entities that influence each other in causal loops</u>. Thus, integrating the DEMATEL framework with econometric modeling can be a potential avenue for future research, where advantages of multiple methods can be leveraged, while countering their shortcomings.

**Acknowledgments**

The authors are thankful to industry experts and academics who have participated in this study. This work would not have been possible without their cooperation and sharing of vision and experiences. We also thank three anonymous reviewers and the editor, whose comments have improved this manuscript significantly.

Xu, X., & Fan, C. K. (2018). Autonomous vehicles, risk perceptions and insurance demand: An individual survey in China. *Transportation Research Part A: Policy and Practice*.

Zhu, Q., Sarkis, J., & Lai, K. H. (2015). Reprint of "Supply chain-based barriers for truck-engine remanufacturing in China". *Transportation Research Part E: Logistics and Transportation Review*, *74*, 94-108.

# Appendix 1: Grey-DEMATEL and Causal Loop Diagramming

***Step 1:*** *Calculate initial direct relation matrices.*

The method begins by collecting the responses of experts in the field. Each expert (*k*) is asked to quantify the influence of factor *i* over factor *j* on a scale with markings: N for "No influence", VL for "Very low influence", L for "Low influence", M for "Medium influence", H for "High influence" and VH for "Very high influence". Let *n* be the number of factors and *K* be the number of experts. Each expert's set of comparisons results in an *n* x *n* matrix, also known as an initial direct relation matrix. With *K* experts, *K* such matrices of size *n* x *n* are obtained.

***Step 2:*** *Compute the average grey-relation matrix.*

Each of the *K* initial relation matrices obtained in Step 1 is first converted into a grey relation matrix using a six-level grey linguistic scale. The mathematical formulation of the grey relation matrix $(X^k)$ is shown in Eq. (1).

$$X^k = \begin{array}{c} \\ B_1 \\ B_2 \\ \vdots \\ B_n \end{array} \begin{array}{cccc} B_1 & B_2 & \cdots & B_n \end{array} \\ \begin{bmatrix} [0,0] & \otimes \tilde{x}_{12}^k & \cdots & \otimes \tilde{x}_{1n}^k \\ \otimes \tilde{x}_{21}^k & [0,0] & \cdots & \otimes \tilde{x}_{2n}^k \\ \vdots & \vdots & \ddots & \vdots \\ \otimes \tilde{x}_{n1}^k & \otimes \tilde{x}_{n2}^k & \cdots & [0,0] \end{bmatrix} \quad (1)$$

where $\otimes \tilde{x}_{ij}^k$ are the grey numbers that indicate the influence of barrier *i* on barrier *j* according to respondent *k*. $B_1$, $B_2$------ $B_n$ indicate the different barriers. All the principal diagonal elements of $X^k$ are set to zero.

$$\otimes \tilde{x}_{ij}^k = \left( \underline{\otimes} \tilde{x}_{ij}^k, \overline{\otimes} \tilde{x}_{ij}^k \right) \quad (2)$$

where $1 \leq k \leq K$; $1 \leq i \leq n$; $1 \leq j \leq n$, and $\underline{\otimes} \tilde{x}_{ij}^k$ and $\overline{\otimes} \tilde{x}_{ij}^k$ represent the lower and upper limits of grey values for respondent *k* in terms of the relationship valuation between factor *i* and factor *j*.

The average grey-relation matrix $A = [\otimes x_{ij}]$ is then obtained from the *K* grey-relation matrices using Equation 3:

$$\otimes x_{ij} = \left( \frac{\sum_k \underline{\otimes} \tilde{x}_{ij}^k}{K}, \frac{\sum_k \overline{\otimes} \tilde{x}_{ij}^k}{K} \right) \quad (3)$$

$$A = [\otimes x_{ij}] \quad (4)$$

***Step 3:*** *Normalize the grey matrix A using the following equations,*

$$\underline{\otimes} \bar{x}_{ij} = (\underline{\otimes} x_{ij} - \min_j \underline{\otimes} x_{ij}) / \Delta_{min}^{max} \tag{5}$$

$$\overline{\otimes} \bar{x}_{ij} = (\overline{\otimes} x_{ij} - \min_j \overline{\otimes} x_{ij}) / \Delta_{min}^{max} \tag{6}$$

Where $\Delta_{min}^{max} = \max_j \overline{\otimes} x_{ij} - \min_j \underline{\otimes} x_{ij}$ (7)

***Step 4:*** *Compute a total normalized crisp value $Y_{ij}$ using the following equation,*

For each element in A, compute,

$$Y_{ij} = \left( \frac{\underline{\otimes}\bar{x}_{ij}(1-\underline{\otimes}\bar{x}_{ij}) + (\overline{\otimes}\bar{x}_{ij} \times \overline{\otimes}\bar{x}_{ij})}{(1 - \underline{\otimes}\bar{x}_{ij} + \overline{\otimes}\bar{x}_{ij})} \right) \tag{8}$$

***Step 5:*** *Determine the final crisp values by the following equations,*

$$z_{ij} = \left( \min_j \underline{\otimes} \bar{x}_{ij} + (Y_{ij} \times \Delta_{min}^{max}) \right) \tag{9}$$

$$Z = [z_{ij}] \tag{10}$$

***Step 6:*** *Obtain a normalized direct crisp relation matrix X using the following equation,*

$$X = \frac{1}{max_{1 \leq i \leq n} \sum_{j=1}^{n} z_{ij}} \times Z \tag{11}$$

***Step 7:*** *Compute the total relation matrix*

The total relation matrix *M* is computed using Equation (12):

$$M = X \times (I - X)^{-1} \tag{12}$$

where *I* represents the identity matrix

***Step 8***: *Calculate row sums $R_i$ and column sums $C_j$*

This is done using Equations 13 and 14:

Sum of columns for row $i$, $R_i = [\sum_{j=1}^{n} m_{ij}]_{n \times 1}$ (13)

Sum of rows for column $j$, $C_j = [\sum_{i=1}^{n} m_{ij}]_{1 \times n}$ (14)

where M= $m_{ij}$, $i, j$=1, 2, --- n

***Step 9:*** *Generate an Influence-Prominence Map using R+C and R–C.*

Each barrier is plotted as a point on a two-dimensional graph – referred to as the Influence Prominence Map (IPM) – using its *R+C* and *R–C* values as its respective x- and y-coordinates. The *x*-axis of the IPM represents "PROMINENCE" and the y-axis stands for "NET INFLUENCE". The "cause" group of barriers will lie above the *y*=0 line on the IPM, while the "effect" group lies below the line. Along the x-axis, barriers that are more towards the right have greater prominence than those towards the left. Essentially, the IPM helps to sort and classify the barriers according to their "PROMINENCE" and "NET INFLUENCE".

***Step 10:*** *Depict the influences using a Causal Loop Diagram (CLD).*

The Causal Loop Diagram (CLD) is central to the systems thinking approach (see Arnold & Wade, 2015; Forrester, 1994; Naweed et al., 2018). A CLD helps depict interrelationships in terms of multiple feedback loops – a chain of influences between factors arranged in a sequence, through which each factor ultimately influences itself (Forrester, 1994; Jia et al., 2019). The CLD predicts the behavior of a system over time better than an approach that views the factors and their interrelationships in isolation (Arnold & Wade, 2015).

The total relations matrix, *M* (see Step 7), provides information on each of the $n$ factors' respective influences on other $(n-1)$ factors, adding up to a total of $n*(n-1)$ influences in the form of distinct $m_{ij}$ values. This can quickly become a large number of influences as the number of factors in the system increases; even in the current study with only 10 factors, this would mean 90 $m_{ij}$ influences. Plotting all the influences can result in a crowding of arrows in the structure drowning the more insightful influences within the weak and insignificant ones. Hence, it has been a practice among DEMATEL users to selectively plot only the relatively stronger influences. For this, a threshold $\theta$ is set and only influences that satisfy $m_{ij} \geq \theta$ are selected. A challenge here is the lack of a clear consensus on how $\theta$ must be set. For example, Rahman and Subramanian (2012) take it to be 0.2, while Ha and Yang (2017) compute $\theta$ as the mean $\mu$ of all $m_{ij}$. In some DEMATEL studies, the standard deviation ($\sigma$) of all $m_{ij}$ is used along with $\mu$, as for example $\theta = \mu + \sigma$ (Bai and Sarkis, 2013), $\theta = \mu + 1.5\sigma$ (Rajesh and Ravi, 2015) and $\theta = \mu + 2\sigma$ (Zhu et al., 2015).